\documentclass[a4paper,11pt]{article}
\usepackage{iftex}
\ifPDFTeX
  \usepackage[T1]{fontenc}
  \usepackage[utf8]{inputenc}
\fi
\usepackage{float}
\usepackage{url}
\usepackage{booktabs}
\usepackage{siunitx}
\usepackage{enumitem}
\usepackage{jheppub}
\usepackage{xcolor}
\usepackage{pgfplots}
\pgfplotsset{compat=1.18}
\usepgfplotslibrary{groupplots}
\definecolor{cfive}{RGB}{0,110,110}   
\definecolor{cten}{RGB}{204,85,0}     
\definecolor{cbar}{RGB}{96,125,139}   
\pgfplotsset{
  simtanka/.style={
    font=\small,
    axis lines*=left,
    axis line style={black!60},
    tick style={black!60},
    tick align=outside,
    grid=major,
    major grid style={black!10},
    label style={font=\small},
    tick label style={font=\footnotesize},
    legend style={draw=none, fill=none, font=\footnotesize},
    legend cell align=left,
    /pgf/number format/1000 sep={},
  },
}
\newcommand{\rainfallplot}[1]{%
\begin{tikzpicture}
\begin{groupplot}[simtanka,
  group style={group size=1 by 2, vertical sep=10pt,
               xticklabels at=edge bottom},
  width=0.92\textwidth, xmin=0.4, xmax=12.6, xtick={1,...,12},
  xticklabels={Jan,Feb,Mar,Apr,May,Jun,Jul,Aug,Sep,Oct,Nov,Dec},
  ybar, /pgf/bar width=0.55, ymin=0, grid=none, ymajorgrids,
  ylabel style={align=center}]
\nextgroupplot[height=0.42\textwidth, ylabel={Rainfall (mm)}]
\addplot[fill=cbar!55, draw=cbar,
  error bars/.cd, y dir=both, y explicit,
  error bar style={black!55}, error mark options={black!55, mark size=2pt}]
  table[x=month, y=mean, y error=std] {#1};
\nextgroupplot[height=0.24\textwidth, ylabel={Wet days}]
\addplot[fill=black!25, draw=black!50] table[x=month, y=wet] {#1};
\end{groupplot}
\end{tikzpicture}}
\usepackage{etoolbox}
\AtBeginEnvironment{table}{\setlength{\abovecaptionskip}{0pt}%
  \setlength{\belowcaptionskip}{8pt}}

\makeatletter
\gdef\@fpheader{}
\makeatother

\title{Modeling Rainwater Harvesting Systems with Covered Storage Tank on A
Smartphone}

\author{Vikram Vyas}
\affiliation{PhysicsInField, New Delhi 110017, India}
\emailAdd{vikram@physicsinfield.org}

\abstract{A model for simulating rainwater harvesting systems with a covered storage tank is described. The model is simple enough to serve as the simulation engine of a smartphone app for designing such systems. The results of the simulation can be used to find the optimum tank size and the reliability of the system in meeting a given demand. This is illustrated by using the model for designing two such systems, one in an urban environment and the other in an arid rural location. The Python scripts used for the simulations are included as ancillary material.}


\begin{document}
\maketitle

\section{Introduction}

Harvesting rainwater and storing it in a covered tank for future use
is an important means of meeting the water needs in many parts of
the world\footnote{For more details see the article \href{https://en.wikipedia.org/wiki/Rainwater_harvesting}{Rainwater Harvesting} in
Wikipedia.}. Such systems are often referred to as rainwater harvesting
systems with covered storage tanks; for brevity we will refer to them
simply as RWHS. These systems are attractive on many counts: they are
relatively simple to build, easy to integrate with other sources of
water, and they minimize the use of energy for pumping water from the
source to the user. The role of RWHS is likely to grow with the
increasing occurrence of drought in regions that historically have not
faced water scarcity \cite{USGCRP:2017aa}.

In exploring the viability of a RWHS for a given location one faces
two critical questions: can it meet the given water demand
reliably, and how much will it cost? The most important factor in
estimating the cost of a RWHS is the size of the tank, so the second
question becomes how big a storage tank will be needed.
As we will see in detail later, the answers to both questions
depend on the amount as well as on the variability of the rainfall,
which together determine the optimum tank size,
that is, the smallest tank beyond which the reliability of the system
no longer improves appreciably. It is expected that due to global
warming this variability will increase in many parts of the world.
This will be reflected in frequent droughts interspersed with floods
\cite{Pendergrass:2017aa}. Taking into account these temporal
fluctuations, along with the spatial variability, requires a detailed
knowledge of the rainfall pattern at the location of the RWHS
\cite{vikSimTanka-95,semaan2020}.

It is here that the advent of smartphones with GPS, and the
availability of large public and commercial rainfall datasets of
increasing resolution and accuracy, has opened up the possibility of
developing a simple but useful model that can be converted into an
app. Such an app can provide a potential user of RWHS with useful
guidelines, such as the optimum tank size for a given demand, the
reliability of the system, and the months when the system will be
under stress and a supplemental source of water may be needed.

The problem of the optimal tank size has been much
studied~\cite{semaan2020}; for examples
see~\cite{liaw2004,campisano2012,fewkes2014}.
In this paper we describe and document one such
model\footnote{The model is called SimTanka and is the simulation
engine for the app RainTank. For details see
\href{https://simtanka.org}{simtanka.org}.}; a precursor of this model was described
in~\cite{vikSimTanka-95}. The outline of the paper is as follows.
In Sec.~\ref{sec:Water-balance-equation} we describe the model. In
Sec.~\ref{sec:Numerical-Experiment-:} we illustrate how the model can
be used to find the optimum tank size for two RWHS in India, one
in Bangalore and one near Jaisalmer. The interpretation and use of the model outputs
are discussed in Sec.~\ref{sec:Interpretation-of-model-outputs}, and
we state our conclusions in the final section,
Sec.~\ref{sec:Conclusion}. The appendix gives the details of the
SimTanka engine and of its Python translation.

\section{The Model}\label{sec:Water-balance-equation}

The question that we are interested in answering is how well a RWHS at
a known location can meet a \textit{given} water demand. One way of
answering the question is to use a simple water balance equation to
evolve the amount of water in the tank, count the number of days on
which the demand cannot be met, and then increase the tank size until
this number no longer decreases appreciably. For this purpose we will
model the RWHS by four parameters: three that characterize the
collecting surface, namely its area $A$, its runoff coefficient $C$
(the fraction of the rainfall on the surface that can be collected)
and its initial abstraction $I_a$ (defined below), together with the
volume $V$ of the tank. The demand is specified by the amount of water
$D_i$ required on each day $i$.

\subsection{Water-balance equation}
Let $r_i$ be the rainfall (mm) on day $i$. A small depth of rainfall
is lost to wetting the surface and filling small depressions before
any runoff occurs; this \emph{initial abstraction} $I_a$ depends on
the catchment surface. The volume of water harvested on day $i$
(m$^3$) is then
\begin{equation}
R_{i}=C\,A\,\max\left(0,\,r_{i}-I_{a}\right)\times10^{-3},
\label{eq:defRunOffCoeff}
\end{equation}
with $A$ in m$^2$. The product $CA$ is the \emph{effective catchment
area}. If $W_{i-1}$ is the water in the tank at the end of the
previous day and $D_i$ the demand (m$^3$) on day $i$, we compute
\begin{equation}
M_{i}=\min\left(W_{i-1}+R_{i},\,V\right)-D_{i},
\label{eq:volumeConstraint}
\end{equation}
and update the tank level as
\begin{equation}
W_{i}=\max\left(M_{i},\,0\right).
\label{eq:evolEq}
\end{equation}
Equation~(\ref{eq:volumeConstraint}) caps the stored water at the
tank volume, any excess being lost as overflow, and
eq.~(\ref{eq:evolEq}) ensures the tank cannot hold a negative amount
of water. Note the order of operations: the day's harvest is added
and the tank capped \emph{before} the day's demand is drawn, so
water that overflows on a given day is not available to meet that
day's demand. This is a conservative choice. Given an initial level
$W_0$, eqs.~(\ref{eq:defRunOffCoeff}--\ref{eq:evolEq}) evolve the
tank level day by day.

The values of the runoff coefficient $C$ and of the initial
abstraction $I_a$ used for different catchment surfaces are listed in
Table~\ref{tab:surfaces}. For roofs the values of $I_a$ follow the
measurements of Farreny et al.~\cite{farreny2011}; the two ground
catchments, which model the prepared and the natural catchment
(\emph{agor}) of a traditional tanka~\cite{vangani1988}, are not
covered by those measurements, and their values are engineering
assumptions taken as the midpoints of the ranges reported in standard
hydrology references~\cite{viessmanlewis2003}.

\begin{table}[tbp]
\centering
\caption{Runoff coefficient $C$ and initial abstraction $I_a$ for each
catchment surface, as implemented in the SimTanka
engine~\cite{simtankarepo}.}\label{tab:surfaces}
\begin{tabular}{l S[table-format=1.2] S[table-format=1.1]}
\toprule
Surface & {$C$} & {$I_a$ (mm)}\\
\midrule
Metal          & 0.95 & 1.0\\
Plastic sheet  & 0.85 & 1.0\\
Concrete       & 0.85 & 1.5\\
Tile / shingle & 0.75 & 2.0\\
Thatch         & 0.30 & 5.0\\
\addlinespace
Paved (prepared) ground$^{*}$ & 0.60 & 2.0\\
Natural sloping ground$^{*}$  & 0.30 & 1.5\\
\bottomrule
\multicolumn{3}{l}{\footnotesize $^{*}$Engineering assumptions; see text.}
\end{tabular}
\end{table}

\subsection{\texorpdfstring{Reliability of the RWHS}{Reliability of the RWHS}}\label{sec:reliability}

Finally we face the question of the rainfall. The approach that
we have taken is the following. Our aim is to estimate the likelihood
of our RWHS meeting the given water demand. For this purpose we evolve the water in the tank using
eqs.~(\ref{eq:defRunOffCoeff}--\ref{eq:evolEq}) over the daily
rainfall record of the \emph{past} five years, starting from an empty tank
($W_{0}=0$). Let $N_{d}$ be the number of days on which the water
demand is non-zero ($D_{i}>0$), and $N_{s}$ the number of those days
on which the system is able to meet the demand, that is, for which
$M_{i}\ge0$. The reliability of the system is then reported as the
percentage
\begin{equation}
P=\frac{N_{s}}{N_{d}}\times100.\label{eq:reliability}
\end{equation}
Starting with an empty tank is the cautious choice: in the first year
the tank has not yet had a chance to fill, so the reliability is
somewhat understated. We use it, as the app does, to find the optimum
tank. Only when we check a tank against later rainfall
(Sec.~\ref{sec:oos}) do we let it fill for one year before we start
counting.

The details of these calculations, together with an explanation of
the actual Swift code that implements them, are available in the
\emph{SimTanka} source repository~\cite{simtankarepo}; see also
Appendix~\ref{sec:engine}.
For a limited evaluation of the future performance, the model
uses the forecast of the rainfall for the next fifteen days in the
water-balance equations~(\ref{eq:defRunOffCoeff}--\ref{eq:evolEq}),
starting from the amount of water in the tank observed by the user,
and reports the number of days on which the demand is met.
In the next section we explicitly check the validity of this
approach for the two sites, and we come back to the validity of its
central assumption, that the recent past is a guide to the near
future, in Sec.~\ref{sec:Interpretation-of-model-outputs}.

\section{\texorpdfstring{Numerical experiments: finding the optimum tank size}{Numerical experiments: finding the optimum tank size}}\label{sec:Numerical-Experiment-:}

We present the results for the use of the model for determining
the optimum tank size for two representative places. Each of these
places is experiencing or is likely to experience water shortage.
We use daily rainfall and temperature obtained from the
\href{https://www.visualcrossing.com/weather-api}{Visual Crossing
Weather API} for 2012--2021, and, for the check of
Sec.~\ref{sec:oos}, for 2011--2025. Each subsection below
states the catchment, roof, and demand inputs used, so the calculation is
reproducible. The optimum tank size for each case is found using the
engine's own convergence-based search: tank size is increased from
$1\,\mathrm{m}^3$ until the reliability computed by
eq.~(\ref{eq:reliability}) stops improving by more than $0.05$ percentage
points over three consecutive steps. In each case we plot the reliability
as a function of storage tank volume using both the daily rainfall for
the past five years and for the past ten years, with the resulting
optimum tank size for each marked on the plot.
In each case we then check the validity of our approach against
later rainfall, which the model has not seen (Sec.~\ref{sec:oos}).

\subsection{Jaisalmer, India}\label{sec:jaisalmer}

As a first example we consider a traditional \emph{tanka}, a covered
underground cistern used in the Thar desert of India, fed not by a
rooftop but by a courtyard or graded ground catchment locally called
an \emph{agor}~\cite{vangani1988}. We consider a small farmer's tanka
used only during the months of July and August, irrigating a small
orchard of twenty ber and pomegranate trees, modelled with the engine's
large-plant category. Supplemental irrigation of these fruit trees from
tanka water is documented by CAZRI~\cite{cazri1998fifty}. We consider an agor whose
surface has been deliberately prepared (lime-plastered, stone, or
cement-treated). Table~\ref{tab:inputs-jaisalmer} states the
catchment and demand inputs used; the catchment area follows
the one concrete figure available in the literature for a related
structure of this kind, a \emph{kund}~\cite{kunds}. The runoff
coefficient $C$ and initial abstraction $I_a$ of the prepared surface
are taken from Table~\ref{tab:surfaces}. The crop coefficients, the
equivalent areas of the plants and the irrigation efficiency used by the
demand model are documented in the SimTanka
repository~\cite{simtankarepo}. The water demand of the
orchard, computed by the demand model from the climate of each window,
is shown in Fig.~\ref{fig:demand-jaisalmer}; over the two months it
amounts to about $59\,\mathrm{m}^3$\footnote{Since the rain falling on the
orchard itself is not subtracted, this demand is an overestimate, and
the optimum tank errs on the safe side: in July and August the summer monsoon
rain supplies much of the trees' water, and the tanka is needed mainly
in the dry spells between rains.}.

The demand is computed from the reference evapotranspiration, which
the engine estimates by the Hargreaves--Samani
method~\cite{hargreaves1985} from the monthly averages of the daily
temperatures, as implemented in the engine's
\texttt{EvapotranspirationCalculator} (Appendix~\ref{sec:engine}). Since these averages are taken over the same window as
the rainfall, the demand changes slightly with the window; at
Jaisalmer by about 1\%.

\begin{figure}[tbp]
\centering
\begin{tikzpicture}
\begin{axis}[simtanka, width=0.6\textwidth, height=0.4\textwidth,
  ybar, /pgf/bar width=18pt, ymin=0, ymax=1250,
  xmin=6.4, xmax=8.6, xtick={7,8}, xticklabels={July, August},
  grid=none, ymajorgrids, ylabel={Demand (L/day)},
  nodes near coords, every node near coord/.append style={font=\scriptsize},
  legend style={at={(0.5,1.03)}, anchor=south, legend columns=2,
    /tikz/every even column/.append style={column sep=1em}}]
\addplot[fill=cfive!75, draw=cfive, area legend] table[x=month, y=w5] {data/Jaisalmer-demand.dat};
\addlegendentry{five-year window, 2017--2021}
\addplot[fill=cten!75, draw=cten, area legend] table[x=month, y=w10] {data/Jaisalmer-demand.dat};
\addlegendentry{ten-year window, 2012--2021}
\end{axis}
\end{tikzpicture}
\caption{Daily water demand of the Jaisalmer orchard of twenty ber and
pomegranate trees, in the two months in which it is irrigated from the
tanka.}\label{fig:demand-jaisalmer}
\end{figure}

\begin{table}[tbp]
\centering
\caption{Case-study inputs for Jaisalmer.}\label{tab:inputs-jaisalmer}
\begin{tabular}{ll}
\toprule
Input & Value\\
\midrule
Latitude, longitude & 26.9157$^\circ$N, 70.9083$^\circ$E\\
Elevation & 225\,m\\
Catchment area & 2000\,m$^2$\\
Surface, runoff coefficient & prepared ground ($C=0.6$)\\
Effective catchment area & 1200\,m$^2$\\
Initial abstraction $I_a$ & 2.0\,mm\\
Demand & orchard of 20 ber and pomegranate trees\\
       & 1005 / 911\,L/day in July / August\\
Active months & July, August\\
\bottomrule
\end{tabular}
\end{table}

\begin{figure}[tbp]
\centering
\rainfallplot{data/Jaisalmer-rain.dat}
\caption{Monthly rainfall pattern for Jaisalmer, India, based on daily
precipitation for 2012--2021. Bars show the monthly mean and whiskers
show one standard deviation; the lower panel shows wet days (daily
precipitation $\protect\geq$ 1 mm). Data: \protect\href{https://www.visualcrossing.com/weather-api}{Visual Crossing Weather API}.
}\label{fig:Rainfall-Jaisalmer}
\end{figure}

\begin{figure}[tbp]
\centering
\begin{tikzpicture}
\begin{axis}[simtanka, width=0.8\textwidth, height=0.5\textwidth,
  xmin=0, xmax=90, ymin=0, ymax=105,
  xlabel={Tank volume (m$^3$)}, ylabel={Reliability (\%)},
  legend pos=north west]
\addplot[cfive, thick] table[x=V, y=P] {data/Jaisalmer-5yr.dat};
\addlegendentry{five-year window, 2017--2021}
\addplot[cten, thick, densely dashed] table[x=V, y=P] {data/Jaisalmer-10yr.dat};
\addlegendentry{ten-year window, 2012--2021}
\draw[black!70, dotted, thick] (axis cs:65,0) -- (axis cs:65,105);
\node[anchor=north east, font=\footnotesize, black!70] at (axis cs:64,60)
  {optimum, 65\,m$^3$};
\end{axis}
\end{tikzpicture}
\caption{Performance of the Jaisalmer tanka, with a prepared agor and
an orchard of twenty fruit trees, as a function of storage tank size.
The dotted vertical line marks the optimum tank size, the same for
both windows.}\label{fig:Jaislmer-Paved}
\end{figure}

The rainfall pattern for Jaisalmer, based on the historical records for
the past ten years, is shown in Fig.~\ref{fig:Rainfall-Jaisalmer}. The
meagerness of the rainfall and the extreme variability is typical of the
arid regions in the world. The performance of the tanka as a function
of the volume of the storage tank is shown in Fig.~\ref{fig:Jaislmer-Paved}.
The optimum tank size is $65\,\mathrm{m}^3$ for both windows, reaching
$100.0\%$ reliability for the five-year window and $92.7\%$ for the
ten-year window. This optimum is close to the total demand for July and
August, about $59\,\mathrm{m}^3$: the tank does not merely bridge dry
days within the monsoon, but stores the season's water and carries it
from one year to the next. In 2021, for example, only 11\,mm of rain fell
in July and August, yet the demand was met in full from water harvested
in 2020.

\subsection{Bangalore, India}\label{sec:bangalore}

Next we consider the case in which RWHS is used for gardening in a
residential house in Bangalore, India, first only during the dry season
and then throughout the year.
Table~\ref{tab:inputs-bangalore}
states the catchment, roof, and demand inputs used for this case.

\begin{table}[tbp]
\centering
\caption{Case-study inputs for Bangalore.}\label{tab:inputs-bangalore}
\begin{tabular}{ll}
\toprule
Input & Value\\
\midrule
Latitude, longitude & 12.9716$^\circ$N, 77.5946$^\circ$E\\
Elevation & 920\,m\\
Roof area, material & 150\,m$^2$, tile ($C=0.75$)\\
Effective catchment area & 112.5\,m$^2$\\
Demand & vegetables 20\,m$^2$,\\
       & 20 small + 10 medium + 2 large plants\\
Active months & December--April (dry season), or all year\\
\bottomrule
\end{tabular}
\end{table}

\begin{figure}[tbp]
\centering
\rainfallplot{data/Bangalore-rain.dat}
\caption{Monthly rainfall pattern for Bangalore, India, based on daily
precipitation for 2012--2021. Bars show the monthly mean and whiskers
show one standard deviation; the lower panel shows wet days (daily
precipitation $\protect\geq$ 1 mm). Data: \protect\href{https://www.visualcrossing.com/weather-api}{Visual Crossing Weather API}.
}\label{fig:Rainfall-Bangalore}
\end{figure}

The rainfall pattern for Bangalore is shown in
Fig.~\ref{fig:Rainfall-Bangalore}. We take the dry season to be
December to April, the months of little rain, for every window. The
demand of the garden in each month is shown in
Fig.~\ref{fig:demand-bangalore}. The
small difference between the windows comes from the temperatures
(Sec.~\ref{sec:jaisalmer}): the ten-year window
2012--2021 was on average slightly warmer than the five-year window
2017--2021, so that the demand is slightly higher. The
performance of the system is shown in
Fig.~\ref{fig:Performance-bangalore-dry}. For the dry season the optimum tank size is
$34\,\mathrm{m}^3$ for the five-year window, reaching a reliability of
$86.6\%$, and $40\,\mathrm{m}^3$ for the ten-year window, reaching
$92.4\%$. As at Jaisalmer, the optimum is close to the demand of the
season, about $40\,\mathrm{m}^3$: the tank has to carry water from the monsoon into the dry
months.

When the same garden is watered throughout the year, the demand is
about $91\,\mathrm{m}^3$ a year for the five-year window and
$94\,\mathrm{m}^3$ for the ten-year window, more than twice that of the
dry season, yet the optimum tank is only $40\,\mathrm{m}^3$, reaching
$93.6\%$, and $56\,\mathrm{m}^3$, reaching $94.5\%$. During the monsoon
the tank is refilled as it is used, so that it has to store little more
than the water of the dry season. Over the ten-year window the
reliability approaches its plateau very slowly: it gains less than one
percentage point between $46$ and $56\,\mathrm{m}^3$, so that the
optimum moves a long way for little change in reliability.

\begin{figure}[tbp]
\centering
\begin{tikzpicture}
\begin{axis}[simtanka, width=0.85\textwidth, height=0.42\textwidth,
  ybar, /pgf/bar width=6pt, ymin=0, ymax=400, xmin=0.4, xmax=12.6,
  xtick={1,...,12}, xticklabels={Jan,Feb,Mar,Apr,May,Jun,Jul,Aug,Sep,Oct,Nov,Dec},
  grid=none, ymajorgrids, ylabel={Demand (L/day)},
  legend style={at={(0.98,0.97)}, anchor=north east, legend columns=1}]
\fill[black!8] (axis cs:0.4,0) rectangle (axis cs:4.5,400);
\fill[black!8] (axis cs:11.5,0) rectangle (axis cs:12.6,400);
\addplot[fill=cfive!75, draw=cfive, area legend] table[x=month, y=w5] {data/Bangalore-demand.dat};
\addlegendentry{five-year window, 2017--2021}
\addplot[fill=cten!75, draw=cten, area legend] table[x=month, y=w10] {data/Bangalore-demand.dat};
\addlegendentry{ten-year window, 2012--2021}
\end{axis}
\end{tikzpicture}
\caption{Daily demand of the Bangalore garden of
Table~\ref{tab:inputs-bangalore} in each month. The shaded months,
December to April, form the dry season.}\label{fig:demand-bangalore}
\end{figure}

\begin{figure}[tbp]
\centering
\begin{tikzpicture}
\begin{groupplot}[simtanka,
  group style={group size=2 by 1, horizontal sep=0.5cm, yticklabels at=edge left},
  width=0.5\textwidth, height=0.45\textwidth,
  xmin=0, xmax=80, ymin=0, ymax=100,
  xlabel={Tank volume (m$^3$)}, title style={font=\small}]
\nextgroupplot[title={Dry season}, ylabel={Reliability (\%)},
  legend to name=bdlegend, legend columns=2,
  legend style={/tikz/every even column/.append style={column sep=1em}}]
\addplot[cfive, thick] table[x=V, y=P] {data/Bangalore-dry-5yr.dat};
\addlegendentry{five-year window, 2017--2021}
\addplot[cten, thick, densely dashed] table[x=V, y=P] {data/Bangalore-dry-10yr.dat};
\addlegendentry{ten-year window, 2012--2021}
\draw[cfive, dotted, thick] (axis cs:34,0) -- (axis cs:34,100);
\draw[cten, dotted, thick] (axis cs:40,0) -- (axis cs:40,100);
\node[anchor=east, font=\footnotesize, cfive] at (axis cs:33.5,40) {34\,m$^3$};
\node[anchor=west, font=\footnotesize, cten] at (axis cs:40.5,40) {40\,m$^3$};
\nextgroupplot[title={Whole year}]
\addplot[cfive, thick] table[x=V, y=P] {data/Bangalore-year-5yr.dat};
\addplot[cten, thick, densely dashed] table[x=V, y=P] {data/Bangalore-year-10yr.dat};
\draw[cfive, dotted, thick] (axis cs:40,0) -- (axis cs:40,100);
\draw[cten, dotted, thick] (axis cs:56,0) -- (axis cs:56,100);
\node[anchor=east, font=\footnotesize, cfive] at (axis cs:39.5,40) {40\,m$^3$};
\node[anchor=west, font=\footnotesize, cten] at (axis cs:56.5,40) {56\,m$^3$};
\end{groupplot}
\end{tikzpicture}

\smallskip
\ref*{bdlegend}
\caption{Performance of the Bangalore RWHS used only during
the dry season, and throughout the year, as a function of the storage tank size. Dotted vertical
lines mark the optimum tank size found for each
window.}\label{fig:Performance-bangalore-dry}
\end{figure}

When the garden is watered all year, the optimum tank of
$40\,\mathrm{m}^3$ meets the demand on 93.6\% of the days over the five
years. Figure~\ref{fig:monthly-bangalore} breaks this down by month.
All the days on which the demand was not met fall between January and
May 2017, the first year of the record, which begins with an empty
tank; from 2018 to 2021 the demand was met on every day. The same holds
for the dry-season garden with its optimum tank of $34\,\mathrm{m}^3$.

\begin{figure}[tbp]
\centering
\begin{tikzpicture}
\begin{axis}[simtanka, width=0.85\textwidth, height=0.38\textwidth,
  ybar, /pgf/bar width=6pt, ymin=0, ymax=34, xmin=0.4, xmax=12.6,
  xtick={1,...,12}, xticklabels={Jan,Feb,Mar,Apr,May,Jun,Jul,Aug,Sep,Oct,Nov,Dec},
  grid=none, ymajorgrids, ylabel={Days demand met},
  legend style={at={(0.5,-0.18)}, anchor=north, legend columns=2,
    /tikz/every even column/.append style={column sep=1em}}]
\addplot[fill=cfive!75, draw=cfive, area legend] table[x=pos, y=avg] {data/Bangalore-year-monthly.dat};
\addlegendentry{average}
\addplot[fill=cten!75, draw=cten, area legend] table[x=pos, y=worst] {data/Bangalore-year-monthly.dat};
\addlegendentry{worst year}
\end{axis}
\end{tikzpicture}
\caption{Monthly reliability of the Bangalore garden watered
throughout the year, with the optimum tank of $40\,\mathrm{m}^3$, over
the five years 2017--2021: the number of days in each month on which
the demand was met in full, on average and in the worst year.}\label{fig:monthly-bangalore}
\end{figure}

In all cases, as expected, there is a value of storage tank
size beyond which the performance remains constant;
Table~\ref{tab:summary} summarizes the optimum tank size and the
reliability found for each. The optimum tank size itself is the same
for the five-year and ten-year windows at Jaisalmer, and larger by
18\% for the ten-year window for the dry-season garden at Bangalore (40\% when
it is watered all year, for the reason given above). The
reliability reached at the plateau differs more between the windows,
mainly because the first year, which starts with an empty tank, counts
for a fifth of a five-year window but only a tenth of a ten-year one.

\begin{table}[tbp]
\centering
\caption{Optimum tank size and reliability at that size.}\label{tab:summary}
\begin{tabular}{l S[table-format=3.0] S[table-format=3.1]
                  S[table-format=3.0] S[table-format=3.1]}
\toprule
 & \multicolumn{2}{c}{Five-year window} & \multicolumn{2}{c}{Ten-year window}\\
\cmidrule(lr){2-3}\cmidrule(l){4-5}
Location & {Tank (m$^3$)} & {Reliability (\%)} & {Tank (m$^3$)} & {Reliability (\%)}\\
\midrule
Jaisalmer             & 65 & 100.0 & 65  & 92.7\\
Bangalore, dry season & {34} & {86.6}  & {40} & {92.4}\\
Bangalore, whole year & {40} & {93.6}  & {56} & {94.5}\\
\bottomrule
\end{tabular}
\end{table}

\subsection{Checking the approach against later rainfall}\label{sec:oos}

For each of the three systems we check the validity of our approach in the following
manner. We first find the optimum tank size using the daily rainfall
for 2011--2020 (ten years), then using that for 2016--2020 (five
years), and finally check the performance of the system with each of
these tanks against the actual daily rainfall for 2021--2025, which
the model has not seen. The sizing windows therefore end in 2020, one
year earlier than in Secs.~\ref{sec:jaisalmer}--\ref{sec:bangalore},
so that none of the test years is used in sizing. The demand is kept
at the value computed for the sizing window, since that is what a user
would have planned with\footnote{Had the demand been recomputed from
the climate of 2021--2025, it would have been 2--6\% lower in every
case; keeping it fixed is therefore slightly conservative.}. The
predicted reliability is that of the app, which starts with an empty
tank. In the test, however, the tank is not new: we let it fill during
2020 and count the days from 1 January 2021. The
results are given in Figs.~\ref{fig:oos-summary} and~\ref{fig:oos}.
Figure~\ref{fig:oos-summary} gives, for each system and sizing window,
the optimum tank, the reliability predicted at that tank over the
sizing window, and the reliability achieved over 2021--2025;
Fig.~\ref{fig:oos} gives the reliability achieved year by year.

{
\begin{figure}[tbp]
\centering
\begin{tikzpicture}
\begin{groupplot}[simtanka,
  group style={group size=3 by 1, horizontal sep=0.4cm, yticklabels at=edge left},
  width=0.37\textwidth, height=0.4\textwidth,
  ybar, /pgf/bar width=9pt, ymin=0, ymax=112, ytick={0,20,...,100},
  symbolic x coords={a,b}, xtick=data, enlarge x limits=0.5,
  x tick label style={font=\scriptsize, align=center},
  grid=none, ymajorgrids, title style={font=\small, align=center},
  nodes near coords={\pgfmathprintnumber[fixed,fixed zerofill,precision=1]{\pgfplotspointmeta}},
  every node near coord/.append style={font=\tiny, rotate=90, anchor=west}]
\nextgroupplot[title={Jaisalmer}, ylabel={Reliability (\%)},
  xticklabels={{sized 2011--20\\65\,m$^3$}, {sized 2016--20\\65\,m$^3$}},
  legend to name=oossumlegend, legend columns=2,
  legend style={/tikz/every even column/.append style={column sep=1em}}]
\addplot[fill=cbar!55, draw=cbar, area legend] coordinates {(a,95.8) (b,93.9)};
\addlegendentry{predicted over the sizing window}
\addplot[fill=cfive!75, draw=cfive, area legend] coordinates {(a,100.0) (b,100.0)};
\addlegendentry{achieved over 2021--2025}
\nextgroupplot[title={Bangalore, dry season},
  xticklabels={{sized 2011--20\\41\,m$^3$}, {sized 2016--20\\32\,m$^3$}}]
\addplot[fill=cbar!55, draw=cbar] coordinates {(a,93.8) (b,86.0)};
\addplot[fill=cfive!75, draw=cfive] coordinates {(a,99.7) (b,93.4)};
\nextgroupplot[title={Bangalore, whole year},
  xticklabels={{sized 2011--20\\56\,m$^3$}, {sized 2016--20\\42\,m$^3$}}]
\addplot[fill=cbar!55, draw=cbar] coordinates {(a,96.0) (b,90.3)};
\addplot[fill=cfive!75, draw=cfive] coordinates {(a,100.0) (b,99.3)};
\end{groupplot}
\end{tikzpicture}

\smallskip
\ref*{oossumlegend}
\caption{Check of the approach against later rainfall: for each
system and sizing window, the optimum tank, the reliability predicted
at that tank over the sizing window, and the reliability achieved with
the rainfall of 2021--2025.}\label{fig:oos-summary}
\end{figure}

\begin{figure}[tbp]
\centering
\begin{tikzpicture}
\begin{groupplot}[simtanka,
  group style={group size=3 by 1, horizontal sep=0.4cm, yticklabels at=edge left},
  width=0.37\textwidth, height=0.4\textwidth,
  ybar, /pgf/bar width=4pt, ymin=0, ymax=112,
  xmin=2020.4, xmax=2025.6, xtick={2021,...,2025},
  x tick label style={font=\scriptsize},
  ytick={0,20,...,100}, grid=none, ymajorgrids,
  title style={font=\small}]
\nextgroupplot[title={Jaisalmer}, ylabel={Reliability (\%)},
  legend to name=ooslegend, legend columns=2,
  legend style={/tikz/every even column/.append style={column sep=1em}}]
\addplot[fill=cten!75, draw=cten, area legend] table[x=year, y=w10] {data/oos-Jaisalmer.dat};
\addlegendentry{sized on 2011--2020}
\addplot[fill=cfive!75, draw=cfive, area legend] table[x=year, y=w5] {data/oos-Jaisalmer.dat};
\addlegendentry{sized on 2016--2020}
\nextgroupplot[title={Bangalore, dry season}]
\addplot[fill=cten!75, draw=cten] table[x=year, y=w10] {data/oos-Bangalore-dry.dat};
\addplot[fill=cfive!75, draw=cfive] table[x=year, y=w5] {data/oos-Bangalore-dry.dat};
\nextgroupplot[title={Bangalore, whole year}]
\addplot[fill=cten!75, draw=cten] table[x=year, y=w10] {data/oos-Bangalore-year.dat};
\addplot[fill=cfive!75, draw=cfive] table[x=year, y=w5] {data/oos-Bangalore-year.dat};
\end{groupplot}
\end{tikzpicture}

\smallskip
\ref*{ooslegend}
\caption{Reliability achieved in each year 2021--2025 by the
tank sized on 2011--2020 and on 2016--2020.}\label{fig:oos}
\end{figure}

\emph{Jaisalmer.} The tank sized on either window met the demand on
every day of 2021--2025, better than predicted. This includes 2021,
the poorest monsoon in the record: the tank carried the water of 2020
through it.

\emph{Bangalore, dry season.} The tank sized on 2011--2020 met the
demand almost every day. The tank sized on 2016--2020 is smaller,
$32\,\mathrm{m}^3$ against $41\,\mathrm{m}^3$. Over the five years it
still did better than predicted, but in 2024 it met the demand on only
83\% of the days (Fig.~\ref{fig:oos}).

\emph{Bangalore, whole year.} The tanks sized on either window met the
demand on every day, or nearly so, better than predicted.

In every case the tank did at least as well as the app predicted. The
app's prediction, which starts with an empty tank, is therefore on the
safe side.
\par}

\section{Interpretation and use of model outputs}\label{sec:Interpretation-of-model-outputs}

The main purpose for developing this model was to facilitate the design,
maintenance and use of a RWHS. Some modelling choices are necessarily
pragmatic. The use of the past five years of daily rainfall records to
estimate future performance is an ad-hoc, but necessary, assumption in
the absence of a dynamical model that can predict future rainfall with
useful precision. The choice of five rather than ten years of daily
rainfall records is a compromise between exposing the system to rare
events and representing interannual variability. The check of
Sec.~\ref{sec:oos} shows that, at least for the systems studied
here, this is not an unreasonable approach.

We now come back to the central assumption of the model, that
the recent past is a useful guide to the near future. The check of
Sec.~\ref{sec:oos} supports it over the period 2021--2025: in every
case the tank sized from the preceding five or ten years of rainfall
did at least as well as predicted. Five years of test, at
two sites, is a short record, and a longer record and more sites would
be needed before the assumption can be regarded as established.

The model is most useful when it is used to answer two sets of practical questions:
\begin{enumerate}
\item Will the reliability of my system improve if I increase the existing size
of the storage tank, and what reliability can I expect from the tank I
can afford? What can I do to increase it: decrease the demand, or
increase the size of the tank? The model answers this by computing the
reliability for a range of tank sizes and finding the plateau beyond
which a larger tank brings no appreciable gain, as in
Figs.~\ref{fig:Jaislmer-Paved} and~\ref{fig:Performance-bangalore-dry}.
Since the cost of the system is set mainly by the size of the tank,
this is the tank worth building. Beyond the optimum size a larger tank
does not help, and the reliability can then be increased only by
reducing the demand; the engine finds the smallest uniform cut in
demand needed to reach a target reliability. The reliability curve also
gives the elasticity of the reliability, which tells how much is lost
with a tank smaller than the optimum. Examples are given below.
\item Given an initial amount of water in the tank, how might the system
meet my water needs over the next fifteen days? Starting
from the amount of water the user observes in the tank, the model runs
the water-balance equations over the rainfall forecast and reports the
days on which the demand will be met. This allows the user to plan
ahead, for example by arranging a supplemental supply or by reducing
use, before the tank runs dry.
\end{enumerate}

To illustrate the first question, consider a user who cannot
afford the optimum tank. At Jaisalmer, CAZRI recommends a tanka of
$50\,\mathrm{m}^3$ for a small nursery of about 200
plants~\cite{goyal2009tanka}; built in masonry, such a tanka cost about
Rs~95{,}000 at 2008 rates, and the cost per litre of the larger tankas
changes little with size, so the optimum tank of $65\,\mathrm{m}^3$
would cost roughly 30\% more. In Bangalore, where rainwater harvesting
has been mandatory since 2009~\cite{bwssb2009}, the regulations ask for a
storage of only 20 litres per square metre of roof area, that is
$3\,\mathrm{m}^3$ for this roof, a tank far too small to carry water
from the monsoon into the dry season; the house considered here has in
fact been built with a tank of $15\,\mathrm{m}^3$. Figure~\ref{fig:smaller}
shows, for the tanks of $50\,\mathrm{m}^3$ at Jaisalmer and
$15\,\mathrm{m}^3$ at Bangalore, watered in the dry season and all year, how the reliability over the five-year
window increases as the demand is reduced uniformly in every month; the
engine reports the smallest such cut, in steps of 5\%, that reaches a
target reliability.

\begin{figure}[tbp]
\centering
\begin{tikzpicture}
\begin{groupplot}[simtanka,
  group style={group size=3 by 1, horizontal sep=0.35cm, yticklabels at=edge left},
  width=0.37\textwidth, height=0.42\textwidth,
  xmin=0, xmax=80, ymin=55, ymax=102, xtick={0,20,...,80},
  xlabel={Cut in demand (\%)}, title style={font=\footnotesize, align=center}]
\nextgroupplot[title={Jaisalmer, 50\,m$^3$\\(optimum 65\,m$^3$)}, ylabel={Reliability (\%)}]
\draw[black!45, densely dashed] (axis cs:0,80) -- (axis cs:80,80);
\draw[black!45, densely dashed] (axis cs:0,90) -- (axis cs:80,90);
\addplot[cten, thick, mark=*, mark size=1.2pt] table[x=cut, y=Jaisalmer] {data/cut-curves-5yr.dat};
\nextgroupplot[title={Bangalore, dry season, 15\,m$^3$\\(optimum 34\,m$^3$)}]
\draw[black!45, densely dashed] (axis cs:0,80) -- (axis cs:80,80);
\draw[black!45, densely dashed] (axis cs:0,90) -- (axis cs:80,90);
\addplot[cfive, thick, mark=square*, mark size=1.2pt] table[x=cut, y=Bangalore-dry] {data/cut-curves-5yr.dat};
\addplot[only marks, mark=*, mark size=2.4pt, cfive] coordinates {(40,84.11) (55,90.33)};
\node[anchor=south east, font=\scriptsize, cfive] at (axis cs:40,84.4) {40\%};
\node[anchor=south east, font=\scriptsize, cfive] at (axis cs:55,90.6) {55\%};
\nextgroupplot[title={Bangalore, whole year, 15\,m$^3$\\(optimum 40\,m$^3$)}]
\draw[black!45, densely dashed] (axis cs:0,80) -- (axis cs:80,80);
\draw[black!45, densely dashed] (axis cs:0,90) -- (axis cs:80,90);
\node[anchor=north east, font=\scriptsize, black!60] at (axis cs:79.5,80) {80\%};
\node[anchor=north east, font=\scriptsize, black!60] at (axis cs:79.5,90) {90\%};
\addplot[cfive!60!black, thick, mark=o, mark size=1.4pt] table[x=cut, y=Bangalore-year] {data/cut-curves-5yr.dat};
\addplot[only marks, mark=*, mark size=2.4pt, cfive!60!black] coordinates {(15,81.04) (40,90.41)};
\node[anchor=north west, font=\scriptsize, cfive!60!black] at (axis cs:15,80.7) {15\%};
\node[anchor=south east, font=\scriptsize, cfive!60!black] at (axis cs:40,90.7) {40\%};
\end{groupplot}
\end{tikzpicture}
\caption{Reliability of tanks smaller than the optimum as the demand
is reduced uniformly in every month, over the five-year window
2017--2021. Filled points mark the
smallest cut that reaches 80\% and 90\%; at Jaisalmer no cut is
needed. The demand is that of Figs.~\ref{fig:demand-jaisalmer}
and~\ref{fig:demand-bangalore}.}\label{fig:smaller}
\end{figure}

At Jaisalmer the smaller tank, about a quarter cheaper, still meets
the demand on 93\% of the days over the five-year window, and no cut is
needed to reach either 80\% or 90\%; over the ten-year window it
reaches 89.2\%, and a cut of 5\% takes it to 90\%. The shortfall falls
in the drier years: in the worst July, in 2019, the demand was met on 16
of its 31 days. In Bangalore the tank of $15\,\mathrm{m}^3$, less than
half the optimum, meets the dry-season demand on 58\% of the days. A cut of
40\% in the demand reaches 80\%, and a cut of 55\% reaches 90\% (the same
cuts over the ten-year window). When the garden is watered all year the
same tank does better, meeting the demand on 77\% of the days, and cuts
of 15\% and 40\% reach 80\% and 90\% (20\% and 45\% over the ten-year
window). The optimum found by the
plateau rule is therefore not a sharp requirement. Where the reliability
approaches its plateau gently, as at Jaisalmer or for the garden
watered all year, a tank well below the
optimum loses little; where it rises steeply up to the plateau, as for
the dry-season garden at Bangalore, it loses much more. We return to
this below (Fig.~\ref{fig:elasticity} and
eq.~(\ref{eq:elasticity})).

Finally, the uniform cut is only a gross indication
of the quantum of reduction needed; how the demand is actually
reduced, and in which months, can be decided only by the user.

Why do the two sites respond so differently to a tank smaller
than the optimum? To see the shape of the reliability curve itself, we
let the tank fill during 2011, count the days of 2012--2021, and scale
the tank by its optimum. Figure~\ref{fig:elasticity}
shows the reliability so obtained for the three systems. A single number
characterises each curve, the elasticity of reliability at half the
optimum tank,
\begin{equation}
E(0.5)=\frac{R(V_{\rm opt})-R(0.5\,V_{\rm opt})}{0.5\,R(V_{\rm opt})},
\label{eq:elasticity}
\end{equation}
the fractional loss of reliability per fractional reduction of the
tank over the second half of the tank, which is the slope of the dashed
lines in Fig.~\ref{fig:elasticity}. It is 0.25 at Jaisalmer and 0.61 for
the dry-season garden at Bangalore, but only 0.23 when the same garden
is watered all year. What decides the sensitivity is therefore when the
water is used rather than the site. When the garden is watered
through the rainy months, the rain refills the tank almost as fast as
the water is used, so even a small tank meets the demand on most days;
a larger tank helps only in the dry spells. A smaller tank therefore
costs little reliability. When the demand falls in a season with almost no rain the
whole season's water must be stored in advance, and the reliability
grows almost in proportion to the tank.

The elasticity tells the user how much the optimum matters. Where it
is high, as for the dry-season garden, the optimum is close to the tank
actually needed. Where it is low, as at Jaisalmer or for the garden
watered all year, the optimum is only an upper limit, and a much
smaller tank loses little reliability. The final choice is therefore
always the user's. In the app, the user is shown the optimum tank and
its reliability, and is asked to enter the size of the tank she has or
plans to build; its reliability is displayed at once. The user can thus
try different sizes and balance reliability against cost in an
informed way.

\begin{figure}[tbp]
\centering
\begin{tikzpicture}
\begin{groupplot}[simtanka,
  group style={group size=3 by 1, horizontal sep=0.35cm},
  width=0.375\textwidth, height=0.36\textwidth,
  xmin=0, xmax=1.2, ymin=0, ymax=103, xtick={0,0.5,1},
  xlabel={Tank / optimum tank},
  title style={font=\footnotesize, align=center}]
\nextgroupplot[title={Jaisalmer, July--August\\$E(0.5)=0.25$}, ylabel={Reliability (\%)},]
\addplot[cfive, thick] table[x expr=\thisrow{V}/65, y=R] {data/warmup-10yr-Jaisalmer.dat};
\addplot[cten, densely dashed, mark=*, mark size=1.6pt] coordinates {(0.5,87.3) (1,100.0)};
\node[anchor=north west, font=\scriptsize, cten] at (axis cs:0.52,87.3) {87.3\%};
\node[anchor=north, font=\scriptsize, cfive] at (axis cs:1,96) {65\,m$^3$};
\nextgroupplot[title={Bangalore, dry season\\$E(0.5)=0.61$}, yticklabels={},]
\addplot[cfive, thick] table[x expr=\thisrow{V}/40, y=R] {data/warmup-10yr-Bangalore-dry.dat};
\addplot[cten, densely dashed, mark=*, mark size=1.6pt] coordinates {(0.5,69.5) (1,100.0)};
\node[anchor=north west, font=\scriptsize, cten] at (axis cs:0.52,69.5) {69.5\%};
\node[anchor=north, font=\scriptsize, cfive] at (axis cs:1,96) {40\,m$^3$};
\nextgroupplot[title={Bangalore, whole year\\$E(0.5)=0.23$}, yticklabels={},]
\addplot[cfive, thick] table[x expr=\thisrow{V}/56, y=R] {data/warmup-10yr-Bangalore-year.dat};
\addplot[cten, densely dashed, mark=*, mark size=1.6pt] coordinates {(0.5,87.4) (1,98.6)};
\node[anchor=north west, font=\scriptsize, cten] at (axis cs:0.52,87.4) {87.4\%};
\node[anchor=north, font=\scriptsize, cfive] at (axis cs:1,94.6) {56\,m$^3$};
\end{groupplot}
\end{tikzpicture}
\caption{Reliability against the tank scaled by its optimum, for
2012--2021 with 2011 as the warm-up year. The dashed line
joins half the optimum tank to the optimum; its slope is $E(0.5)$. The optimum tank is marked in each panel.}\label{fig:elasticity}
\end{figure}

{The results of Sec.~\ref{sec:Numerical-Experiment-:} also raise the
question of how stable the optimum tank is between the five-year and
ten-year windows.

\emph{Jaisalmer.} The optimum tank is $65\,\mathrm{m}^3$ for both
windows.

\emph{Bangalore, dry season.} The optimum tank changes from 34 to
$40\,\mathrm{m}^3$ between the windows.

\emph{Bangalore, whole year.} The optimum tank moves from 40 to
$56\,\mathrm{m}^3$, but over the ten-year window the reliability rises
by less than one percentage point beyond $46\,\mathrm{m}^3$. Near its
plateau this curve is so flat that the exact optimum matters little, as
its small elasticity $E(0.5)$ also shows.

Whether the optimum stays as stable over longer records and at other
sites is worth exploring.
\par}

\subsection{\texorpdfstring{Sensitivity to the source of rainfall data}{Sensitivity to the source of rainfall data}}\label{sec:datasource}

{
The results of this paper were computed with daily data from Visual
Crossing. To see how much they depend on the source, the whole analysis
was repeated, with the same scripts and the same dry season, on the
ERA5 reanalysis~\cite{era5} for the same coordinates, 2011--2025, as
served by Open-Meteo~\cite{openmeteo}\footnote{By default Open-Meteo serves a
``best match'' that combines ERA5 with the ECMWF operational forecast
model from 2017 onwards, so that its record is not consistent in time;
with that record the optimum tank at Jaisalmer falls from 65 to about
$50\,\mathrm{m}^3$. The ERA5 record alone is obtained with the option
\path{models=era5}, which the scripts provide. So that the results of
this section can be reproduced exactly, and without any key, the ERA5
data used here are included in the ancillary files, in the folder
\path{data-era5}.}. Figure~\ref{fig:datasource}
compares the optimum tanks.

\begin{figure}[tbp]
\centering
\begin{tikzpicture}
\begin{groupplot}[simtanka,
  group style={group size=3 by 1, horizontal sep=0.4cm, yticklabels at=edge left},
  width=0.37\textwidth, height=0.4\textwidth,
  ybar, /pgf/bar width=10pt, ymin=0, ymax=80, grid=none, ymajorgrids,
  symbolic x coords={a,b}, xtick=data, xticklabels={five-year, ten-year},
  enlarge x limits=0.5, x tick label style={font=\scriptsize},
  nodes near coords, every node near coord/.append style={font=\tiny},
  title style={font=\small}]
\nextgroupplot[title={Jaisalmer}, ylabel={Optimum tank (m$^3$)},
  legend to name=dslegend, legend columns=2,
  legend style={/tikz/every even column/.append style={column sep=1em}}]
\addplot[fill=cfive!75, draw=cfive, area legend] coordinates {(a,65) (b,65)};
\addlegendentry{Visual Crossing}
\addplot[fill=cten!75, draw=cten, area legend] coordinates {(a,62) (b,62)};
\addlegendentry{ERA5}
\nextgroupplot[title={Bangalore, dry season}]
\addplot[fill=cfive!75, draw=cfive] coordinates {(a,34) (b,40)};
\addplot[fill=cten!75, draw=cten] coordinates {(a,36) (b,38)};
\nextgroupplot[title={Bangalore, whole year}]
\addplot[fill=cfive!75, draw=cfive] coordinates {(a,40) (b,56)};
\addplot[fill=cten!75, draw=cten] coordinates {(a,40) (b,47)};
\end{groupplot}
\end{tikzpicture}

\smallskip
\ref*{dslegend}
\caption{Optimum tank for the five-year and ten-year windows,
found with the two sources of daily data.}\label{fig:datasource}
\end{figure}

\begin{figure}[tbp]
\centering
\begin{tikzpicture}
\begin{groupplot}[simtanka,
  group style={group size=2 by 1, horizontal sep=0.5cm, yticklabels at=edge left},
  width=0.45\textwidth, height=0.42\textwidth,
  ybar, /pgf/bar width=14pt, ymin=0, ymax=110, ytick={0,20,...,100},
  symbolic x coords={p,t}, xtick=data, enlarge x limits=0.6,
  xticklabels={{tested on\\Visual Crossing}, {tested on\\ERA5}},
  x tick label style={font=\scriptsize, align=center},
  grid=none, ymajorgrids, title style={font=\small, align=center},
  nodes near coords={\pgfmathprintnumber[fixed,fixed zerofill,precision=1]{\pgfplotspointmeta}},
  every node near coord/.append style={font=\tiny}]
\nextgroupplot[title={Bangalore, dry season\\tanks of 32 and 36\,m$^3$},
  ylabel={Reliability 2021--2025 (\%)},
  legend to name=ctlegend, legend columns=2,
  legend style={/tikz/every even column/.append style={column sep=1em}}]
\addplot[fill=cfive!75, draw=cfive, area legend] coordinates {(p,93.4) (t,93.0)};
\addlegendentry{tank sized on Visual Crossing}
\addplot[fill=cten!75, draw=cten, area legend] coordinates {(p,99.1) (t,99.9)};
\addlegendentry{tank sized on ERA5}
\nextgroupplot[title={Bangalore, whole year\\tanks of 42 and 40\,m$^3$}]
\addplot[fill=cfive!75, draw=cfive] coordinates {(p,99.3) (t,94.2)};
\addplot[fill=cten!75, draw=cten] coordinates {(p,99.5) (t,95.2)};
\end{groupplot}
\end{tikzpicture}

\smallskip
\ref*{ctlegend}
\caption{The tanks sized on 2016--2020 from each record, tested against
the rainfall of 2021--2025 from each record, for the dry-season
garden and the garden watered all year.}\label{fig:crosstest}
\end{figure}

\emph{Jaisalmer.} The two records are almost the same, day by day, so
the results hardly differ. The optimum tank is $65\,\mathrm{m}^3$ with
Visual Crossing and $62\,\mathrm{m}^3$ with ERA5
(Fig.~\ref{fig:datasource}).

\emph{Bangalore.} Here the two records are quite different. ERA5 has
about a fifth less rain than Visual Crossing, spread over more days.
This is because ERA5 does not record the rain at one place: it
estimates the average rainfall over an area some 30\,km
across~\cite{era5}, and averaging over an area spreads the rain more
thinly over more days.

\emph{Bangalore, dry season.} The optimum tank found from the two
records differs by at most $2\,\mathrm{m}^3$
(Fig.~\ref{fig:datasource}). In the check of Sec.~\ref{sec:oos}, the
tank sized on the five years 2016--2020 is $32\,\mathrm{m}^3$ with
Visual Crossing data and $36\,\mathrm{m}^3$ with ERA5 data. Does the
result of the check depend on the tank, or on the rainfall used to
test it? To find out, we tested both tanks against the rainfall of both
records (Fig.~\ref{fig:crosstest}). Each tank did about as well with
either rainfall, and both did better than predicted; the larger tank
met the demand on more days. So it is the size of the tank that
matters, not the record used to test it.

\emph{Bangalore, whole year.} Here the reliability rises only slowly as
the tank grows beyond about $40\,\mathrm{m}^3$, so a small difference
in the rainfall can move the optimum a long way. For the ten-year
window it is $56\,\mathrm{m}^3$ with Visual Crossing and
$47\,\mathrm{m}^3$ with ERA5 (Fig.~\ref{fig:datasource}). In the check of Sec.~\ref{sec:oos}, the two tanks performed about
equally well (Fig.~\ref{fig:crosstest}).

\emph{In short,} the two records give approximately the same tank, and
that is what matters to the user. We comment further on this in the
Conclusion (Sec.~\ref{sec:Conclusion}).
\par}

\section{Conclusion}\label{sec:Conclusion}

RWHS are useful means of meeting the water needs in the water scarce
regions of the world. Their attractiveness is tempered by their potential
unreliability as they depend on the vagaries of the daily rainfall.
The aim of the model described here is to provide a practical
guideline for designing a RWHS, in terms of the optimal tank size and
expected reliability.
For the examples studied here, the model gives a clear answer to how
large the tank should be. Just as important, it gives the reliability
of any other tank, so that the user can weigh reliability against
cost. When these tanks were tested against the
rainfall of 2021--2025, which was not used to size them, they did at
least as well as predicted. Two global datasets that differ a good
deal at a given place also give approximately the same tank size
(Sec.~\ref{sec:datasource}).

A natural question is whether the probabilistic model described
here can be replaced, or supplemented, by machine learning models.
Although earlier assessments of machine learning approaches to rainfall
prediction were rather cautious~\cite{hussein2022rainfall}, recent
advances in deep learning and AI-based weather forecasting have
substantially improved their capabilities~\cite{lam2023graphcast}.
Nevertheless, predicting daily rainfall at a specific location,
particularly over the extended periods relevant to rainwater harvesting
systems, remains a significant challenge. A promising possibility is to bring together local rainfall
measurements and a real-time record of the water level in the tank.
From these, the parameters of the model, such as the runoff coefficient
and the demand, can be estimated for the actual system, and AI-based
methods could use the same data to make better short-term predictions
of how the system will perform.

In this context, the development of inexpensive automated rain gauges
and tank water-level sensors, capable of communicating with smartphones
through open protocols, would be of considerable benefit. Recent
developments in low-cost wireless hydrological monitoring systems
demonstrate the technical feasibility of such an
approach~\cite{abdelal2021}, although questions of calibration,
reliability, and long-term maintenance remain important~\cite{zhu2023lowcost}.

Box and Draper remarked that ``essentially, all models are wrong, but
some are useful''~\cite{box1987}. A real RWHS can never be
described fully by a model with only four parameters. The real
question, then, is: is this model useful? SimTanka was developed
to help a user balance two conflicting aims: to catch as much rain as
possible, reliably, and to keep the cost low. The results of this paper
indicate that, to this extent, the answer is yes.

\section*{Use of AI}

The major use of AI was in translating the SimTanka engine from pure
Swift into the Python scripts that are included as ancillary material.
The appendix describing the engine was written by an AI model under my supervision. In
addition, AI was used to improve the figures and tables, to find
relevant references, and in preparing the \LaTeX{} document.
I take full responsibility for the content of the manuscript.

\acknowledgments

The development of SimTanka was started when I was working with the
\href{https://ajitfoundation.net}{Ajit Foundation}, and I would like to
thank the Foundation for its support and encouragement. I am grateful
to Janaki Abraham for her continuing support for this project, and to
Visual Crossing Weather for generously providing the historical
rainfall data used to develop and refine the SimTanka model. I also
thank Open-Meteo for making its historical weather data freely
available; the data used in Sec.~\ref{sec:datasource} are from
Open-Meteo.com (CC BY 4.0) and contain modified Copernicus Climate
Change Service information.

\appendix

\section{The SimTanka engine}\label{sec:engine}

The model of Sec.~\ref{sec:Water-balance-equation} is implemented in
\emph{SimTanka}, an open-source engine written in Swift and released
under the MIT licence~\cite{simtankarepo}. It is distributed as a Swift
package (\texttt{RainTankEngine}) that depends only on the Foundation
library and builds on macOS, iOS and Linux. The engine is purely a
computation library: every function takes plain values as arguments
and returns plain values, with no networking, storage or user
interface. In particular it fetches no weather data. The user supplies
the daily rainfall and, for the demand model, daily temperatures,
latitude and elevation, from whatever source they choose. Each
equation and constant is documented, with its source, in the
repository, so that the model can be reimplemented in other languages.

\subsection{What the engine computes}

\begin{itemize}[nosep]
\item \emph{Historical reliability}: $P$ of eq.~(\ref{eq:reliability})
  over the most recent five years of rainfall.
\item \emph{Optimum tank size}: the plateau search of
  Sec.~\ref{sec:Numerical-Experiment-:}.
\item \emph{Monthly reliability}: the same continuous simulation,
  broken down by calendar month and reporting both the average and
  the worst year, which shows \emph{when} in the year the system
  tends to fail.
\item \emph{Uniform demand reduction}: the smallest uniform cut in
  demand, searched in steps of 5\% up to 80\%, needed to reach a
  target reliability with a given tank.
\item \emph{Forecast}: the fifteen-day projection of
  Sec.~\ref{sec:reliability}.
\item \emph{Water demand}: the monthly demand profile from household
  and garden sub-models, documented in the repository.
\end{itemize}
The illustrations in Sec.~\ref{sec:Numerical-Experiment-:} use the
first three.

\subsection{A minimal example}

The following computes the reliability of a system and its optimum
tank size. The catchment is described by its effective area $CA$ and
initial abstraction (Table~\ref{tab:surfaces}), the demand by twelve
monthly values in litres per day, and the rainfall by daily records
keyed by year.
\begin{verbatim}
import RainTankEngine

let surface = RoofMaterial.tile
let config = SystemConfiguration(
    effectiveCatchmentArea: 40.0 * surface.runoffCoefficient,
    tankVolume: 5.0,                       // m^3
    monthlyDemands: demandLitresPerDay,    // 12 values
    initialAbstractionMM: surface.initialAbstractionMM)

let engine = SimTankaEngine()
let result = try engine.calculateHistoricalReliability(
    config: config, rainfallData: rainfall)
let optimum = try engine.findOptimalTankSize(
    config: config, targetReliability: 1.5, rainfallData: rainfall)
\end{verbatim}
A target above $1$ (that is, above 100\%) can never be reached, so the
search stops only when the reliability curve levels off, as in
Sec.~\ref{sec:Numerical-Experiment-:}.

\subsection{Python translation and reproducibility}

For this paper the engine and the demand model were translated line
by line into Python. The one deliberate change is that the length of
the window is a parameter rather than fixed at five years. This
applies both to the rainfall used in the simulation and to the
climate statistics from which the garden demand is computed, so a
ten-year run differs from a five-year run in its demand as well as in
its rainfall. The translation was checked against the Swift engine by running
both on identical inputs, for the five-year window 2017--2021 (the one
fixed in the Swift implementation) and for the systems of this paper:
the Jaisalmer orchard, the Bangalore garden watered in the dry season,
and the same garden watered all year. For all three, the reliabilities
agree exactly at every tank size from 1 to 100~m$^3$, the optimum tank
sizes are identical, the monthly reliability
and the uniform cut in demand agree exactly, and the monthly demands agree to
within floating-point rounding (relative differences below
$10^{-14}$). The check against later rainfall (Sec.~\ref{sec:oos}) and the
elasticity of Fig.~\ref{fig:elasticity} use the same translated water
balance with a warm-up year, which is not part of the Swift engine.

The ancillary files of this paper contain the Python translation of
the engine and of the demand model (files \path{simtanka_engine.py},
\path{water_demand.py} and \path{monthly_reliability.py}), the
definitions of the two systems (\path{cases.py}), a single script
(\path{paper_results.py}) that reproduces every figure and table of
Secs.~\ref{sec:Numerical-Experiment-:}
and~\ref{sec:Interpretation-of-model-outputs} from the daily weather, a
module that reads and checks the weather files
(\path{rainfall_input.py}), and the script used to download the weather
data (\path{fetch_data.py}). The same files are in the folder
\path{paper/arxiv-v2.2} of release v2.3 of the SimTanka
repository~\cite{simtankarepo}, together with the agreement check
described above (folder \path{parity}). The scripts read the daily rainfall (and, for the demand model,
temperature) from a simple text file in a standard format, so that any
source can be used, including a local rain gauge. Such records are
freely available from public sources, and the included module downloads
them directly, without any key, from the Open-Meteo historical weather
service~\cite{openmeteo} and from NASA POWER~\cite{nasapower}, both of
which provide daily rainfall and temperature worldwide; from
Open-Meteo it can fetch the ERA5 record alone, as used in
Sec.~\ref{sec:datasource}, and the script \texttt{cross\_test.py}
repeats the cross-check of that section. Other free
sources are CHIRPS~\cite{funk2015chirps}, which provides daily rainfall
between latitudes 50$^\circ$S and 50$^\circ$N, and, for India, the
gridded daily rainfall and temperature data of the India Meteorological
Department~\cite{pai2014imd}. The data used in this paper, for
2011--2025, were obtained from the Visual Crossing Timeline API, whose
licence does not allow them to be distributed.

\end{document}